\documentclass[11pt,a4paper]{article}

\usepackage[margin=1in]{geometry}
\usepackage{amsmath,amssymb,amsthm}
\usepackage{booktabs}
\usepackage{graphicx}
\usepackage{hyperref}
\usepackage{natbib}
\usepackage{tikz}
\usetikzlibrary{arrows.meta,positioning,fit,calc}
\usepackage{xcolor}
\usepackage{microtype}
\usepackage{placeins}
\usepackage{enumitem}

\title{When Should Deployed BCI Decoders Adapt?\\
A Resource-Priced Ladder for Frozen EEG Trunks}
\author{
  Sergey Musienko\\
  \texttt{sergey@nimbusbci.com}
}
\date{}

\begin{document}
\maketitle

\begin{abstract}
\emph{When} should a deployed MI decoder spend adaptation resources---not
only \emph{how} to adapt under shift?
We study a thin control layer outside a frozen encoder: a resource ladder
(factory \(\to\) Bayesian head \(\to\) affine map, with full fine-tune as
an escape hatch) governed by regime-aware gating under a utility that
prices accuracy, fit time, and label cost. The head is Bayesian in the
conjugate sense: a Normal--Inverse--Wishart prior over the class
conditional Gaussians in embedding space, updated in closed form (no
gradients) as labelled trials arrive. Full estimators, constants, and
the reanalysis code are in the companion
paper~\citep{musienko2026personalizer} and the ancillary bundle of
this submission.
\textbf{Thesis:} adaptation should be gated and capped, not always-on and
not blindly fine-tuned.
\textbf{Evidence:} (i)~utility-max gating over-adapts on clean data that
already carries session drift; a clean-referenced gate restores low
false-entry rates (e.g.\ dropping the adaptation rate from \(0.56\) to
\(0.11\) on field-like sessions, with the cascade staying on the factory
decoder for \(\approx\!87\%\) of clean cells).
(ii)~Under severe synthetic stress, a lightweight LDA head matches or
exceeds the frozen trunk's accuracy across five encoder architectures
(\(n{=}155\) subject\(\times\)trunk cells, pooled across shared
subjects), at \(\sim\!23\times\) lower fit time than full warm-start
fine-tuning. Always-on baselines favor capping the default below full
fine-tune.
(iii)~In two no-perturbation BNCI session checks the factory model
transferred without intervention (no reliable head gain), so the
control layer's value under natural drift is cost-efficiency and
clean-safety rather than demonstrated accuracy gains.
(iv)~Adaptation severity does not follow a universal ordinal ladder across
datasets, highlighting the need for dynamic gating over fixed adaptation
policies.
\end{abstract}

\section{Introduction}
\label{sec:intro}

Deep convolutional encoders such as EEGNet~\citep{lawhern2018eegnet} and
related architectures~\citep{schirrmeister2017deep} are strong MI
decoders, but session and subject shift remain central failure
modes~\citep{lotte2018review,jayaram2016transfer}.
The same control question applies as encoders scale toward
foundation-model size; the companion paper extends the Personalizer
API in that direction~\citep{musienko2026personalizer}, and the gate +
ladder here are trunk-agnostic by construction, but our evidence is on
task-trained architectures.
The default engineering reflex is to fine-tune---or always adapt---whenever
performance might drop.
However, in deployment the scarcer resource is often the
\emph{decision} to spend labels and fit time, not the mere existence of an
adapter.

We ask: \emph{when and how should a deployed BCI system spend adaptation
resources?}
We study a thin control layer outside a frozen trunk:
\[
x \;\xrightarrow{\;f_\theta\;}\; z \;\xrightarrow{\;\text{head}\;}\; \hat{y},
\]
with optional escalation to a supervised affine map on \(z\) and, as an
escape hatch, full warm-start fine-tuning.
The companion Personalizer paper~\citep{musienko2026personalizer} ships
the trunk-agnostic API that this control layer governs; this paper's
contribution is the \emph{resource-aware ladder} and \emph{regime-aware
gate} that decide whether that surface should spend labels and fit time.
The control layer itself (gate, ladder, utility) is self-contained and
does not require reading the companion.

Two stress roles matter and should not be conflated:
\emph{synthetic channel mixing} provides a controlled severe cell for cost
comparisons; \emph{natural session drift} (especially the online-feedback
Kumar2024 corpus introduced below) stresses the gate that decides whether
to leave the factory decoder.

Empirical questions:
\begin{enumerate}[leftmargin=1.4em,itemsep=0.2em]
\item When should the system leave the factory decoder when ``clean'' is
not factory-clean?
\item How much fine-tune gain can a cheap head capture, and should the
default escalate to full fine-tune under a shared utility metric?
\item Does an ordinal severity ladder transfer across trunks and datasets?
\end{enumerate}

\paragraph{Main takeaway and contributions.}
Adaptation should be gated, not always-on.
A clean-referenced gate plus a utility-priced ladder capped below full
fine-tuning is a stronger product default than constant adaptation.
Our goal is to control the shared Personalizer surface efficiently,
bounding our claims to cost-efficiency and recovery under severe stress,
rather than proposing a universally superior decoder.
Specifically, our contributions are:
\begin{enumerate}[leftmargin=1.4em,itemsep=0.2em]
\item A \textbf{clean-referenced mean-shift gate} that prevents false entry
into head adaptation when data already carries session drift, effectively
restoring low adaptation rates where a utility-max gate overfires
(Section~\ref{sec:cq}).
\item \textbf{Cost-efficient recovery under severe stress}: evidence that a
lightweight LDA head recovers accuracy under severe stress across five
encoder architectures, achieving this at \(\sim\!23\times\) lower fit
time than full fine-tuning (Section~\ref{sec:cost}).
\item \textbf{Utility-driven capping}: a robust argument for capping the
product default below full fine-tuning, evaluated across a varying
utility-weight grid of fit-time and label costs
(Sections~\ref{sec:skipl3}--\ref{sec:usens}).
\item \textbf{Boundary mapping}: we show there is no universal ordinal
severity\(\to\)level policy across trunks and datasets; rather,
factory-when-clean is the most transferable soft rule
(Section~\ref{sec:transfer}).
\end{enumerate}

\paragraph{Related work and positioning.}
Session-to-session adaptation and transfer learning are long-standing
themes in BCI~\citep{lotte2018review,jayaram2016transfer}. Three recent
lines are closest. \emph{Test-time adaptation} updates the decoder as
unlabeled trials arrive: T-TIME~\citep{li2024ttime} maintains an
information-maximization ensemble with a fast unsupervised learner per
class; source-free variants adapt the backbone
online~\citep{wimpff2024calibration}; and LOTL~\citep{wu2026lotl}
applies closed-form, one-pass local transfer to each arriving sample. \emph{Continual supervised
fine-tuning}: EDAPT~\citep{haxel2025edapt} fine-tunes continually
across nine datasets and three tasks toward calibration-free
operation. These lines advance \emph{how} to adapt cheaply without
labels (T-TIME, source-free TTA) or \emph{how far} continual
supervision travels (EDAPT); none of them, per their reports, treats
the \emph{decision to adapt} as a first-class, cost-priced control
variable. That is our contribution and its boundary: an explicit
utility over accuracy, fit time, and labels; a clean-referenced trigger
that controls false \emph{entry} into adaptation; and a capped ladder
over a deliberately frozen trunk
(Table~\ref{tab:related}). The compared families are selected as the
closest recent representatives of ensemble, source-free, one-pass, and
continual adaptation---an illustrative comparison, not a survey.

\begin{table}[t]
\centering
\caption{Positioning against the closest adaptation families. ``Decides
whether to adapt'' = an explicit gate on \emph{entry} into adaptation;
``prices costs'' = fit time and label burden enter the policy's
objective. Among the families compared here, this work is the one
that freezes the trunk by design and prices the decision---a
positioning, not a field survey. $^{\mathrm{a}}$T-TIME replaces the
decoder with a lightweight ensemble; no comparable frozen trunk is
retained.}
\label{tab:related}
\small
\resizebox{\textwidth}{!}{%
\begin{tabular}{lccccc}
\toprule
 & T-TIME & Source-free & LOTL & EDAPT & \textbf{This} \\
 & \citep{li2024ttime} & TTA \citep{wimpff2024calibration} & \citep{wu2026lotl} & \citep{haxel2025edapt} & \textbf{work} \\
\midrule
Supervised updates & no & no & no & yes & \textbf{gated} \\
Trunk frozen by design & no$^{\mathrm{a}}$ & no & no & no & \textbf{yes} \\
Decides whether to adapt & no & no & no & no & \textbf{yes} \\
Prices fit time + labels & no & no & no & no & \textbf{yes} \\
Rung selection online & per-trial & per-trial & one-pass & continual & \textbf{cascade} \\
\bottomrule
\end{tabular}%
}
\end{table}

\section{Setup}
\label{sec:setup}

\subsection{End-to-end pipeline}
\label{sec:pipeline}

\begin{figure}[t]
\centering
\includegraphics[width=0.98\linewidth]{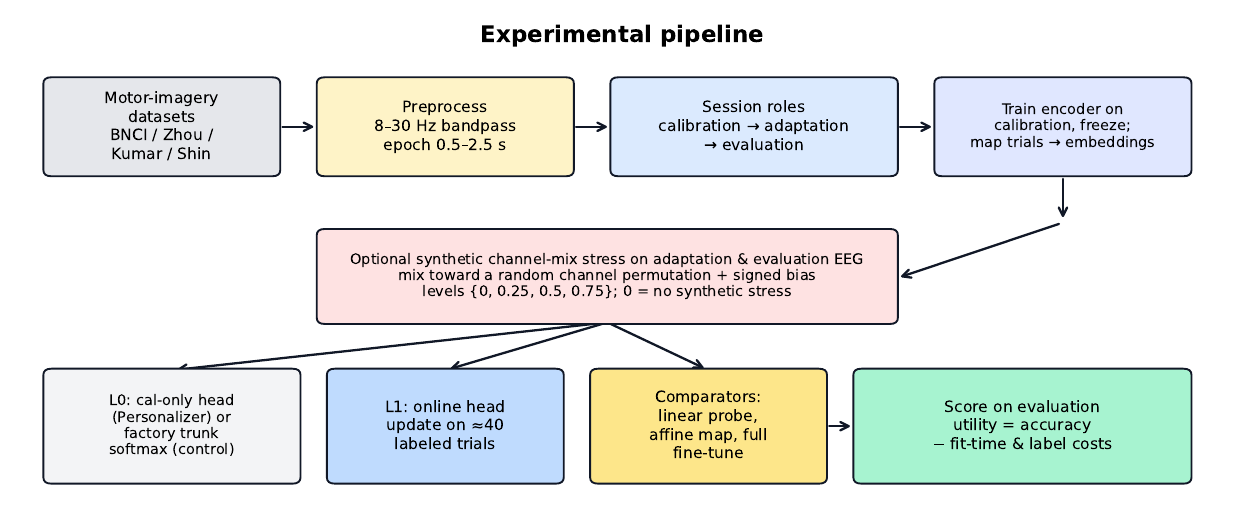}
\caption{Experimental data flow.
Motor-imagery epochs \(\to\) bandpass + epoch window \(\to\) calibration /
adaptation / evaluation sessions \(\to\) train encoder on calibration and
freeze \(\to\) optional synthetic channel-mix stress on adaptation and
evaluation EEG \(\to\) embeddings \(\to\) ladder arms \(\to\) score
under~\eqref{eq:U}.}
\label{fig:pipeline}
\end{figure}

Figure~\ref{fig:pipeline} summarizes the shared experimental data flow
(also shared with the companion Personalizer study).

\paragraph{Datasets and session geometry.}
All ladder experiments use a three-session split: calibration = first
session, adaptation = middle session(s), evaluation = last session
(public MOABB corpora~\citep{moabb2021}).
Datasets: BNCI2014-004~\citep{tangermann2012review} (9 subjects),
Zhou2016~\citep{zhou2016complete} (4), Shin2017A~\citep{shin2017open}
(subjects 1--9; short adaptation session, often \(n{=}20\) labeled trials), and
Kumar2024~\citep{kumar2024online} (18 subjects; sessions after the first
are online with feedback---field-like even without synthetic stress).

\paragraph{Preprocessing.}
Shared motor-imagery filter bank: bandpass \textbf{8--30\,Hz}, post-cue
window \([0.5, 2.5]\)\,s for classical trunks (fair cross-trunk band).
Optional foundation runs may use 1--40\,Hz under the same window; we do not
mix bands inside a classical cost cell.

\paragraph{Trunk then freeze.}
Per subject, the trunk (EEGNet in the cost bake-off; five architectures in
the multi-trunk replication of Section~\ref{sec:cost}) is trained on clean
calibration trials and frozen.
Downstream arms consume frozen-encoder embeddings unless L3 fine-tunes the
trunk.

\paragraph{Adaptation levels.}
\begin{description}[style=nextline,leftmargin=1.2em,font=\normalfont\itshape]
\item[L0 --- factory] Trunk softmax of the frozen encoder; no Personalizer
head and no online adaptation.
\item[L1 --- Bayesian head] The companion paper's \emph{Personalizer}
head (conjugate LDA with shared covariance), on frozen embeddings,
updated online as labeled stream trials arrive.
\item[L2 --- affine embeddings] Supervised affine map \(Z'=ZA+b\), then
head update.
\item[L3 --- full fine-tune] Warm-start fine-tune of the trunk (EEGNet
path in our bake-off).
\end{description}
\paragraph{Notation.}
\emph{Factory} / \emph{frozen trunk softmax} means the pretrained trunk's own
classifier with encoder weights fixed.
\emph{Frozen embeddings} means the same fixed encoder, while a head or affine
map may still adapt on top of it.
We select affine adaptation as L2 because it is the
\emph{smallest supervised transformation} that can correct geometry-level
embedding drift (scale, rotation, and translation in \(Z\)) without
modifying encoder weights or semantics.
Richer alternatives (distribution alignment, Riemannian methods, adapters,
LoRA) are compatible with the same ladder idea; comparing them is outside
our scope.
L2 is a product ceiling under the utility below, not a claim that affine
is the uniquely optimal mid-tier method.

\paragraph{Utility.}
Unless noted, policies share a single utility score \(U\) that trades
classification accuracy against adaptation fit time and labeled budget:
\begin{equation}
\label{eq:U}
U = \mathrm{acc}
  - \lambda_w \cdot t_{\mathrm{adapt}}
  - \mu_n \cdot (n_{\mathrm{adapt}} / N_{\max}),
\end{equation}
where \(\mathrm{acc}\) is evaluation accuracy, \(t_{\mathrm{adapt}}\) is
adaptation fit time in seconds in our protocol (not end-to-end latency),
and \(n_{\mathrm{adapt}}\) is the number of labeled stream trials spent
(\(n_{\mathrm{adapt}}=0\) at L0; \(n_{\mathrm{adapt}}=n\) otherwise).
Each second of fit time penalizes \(U\) by \(\lambda_w\) (in
accuracy-equivalent units), and each block of \(N_{\max}\) labeled trials
penalizes \(U\) by \(\mu_n\).
For our primary cells we evaluate a budget of \(n{=}40\) against a scale
\(N_{\max}{=}80\), fixing the wall-time penalty \(\lambda_w{=}0.02\) and
the label penalty \(\mu_n{=}0.05\). The robustness of the recommended
policy to these specific weights is analyzed in
Section~\ref{sec:usens}.

\paragraph{Stress.}
\begin{sloppypar}
Ladder and cost cells apply a synthetic channel-mix perturbation at
levels \(\{0,0.25,0.5,0.75\}\) unless stated: a channel-permutation mix
plus signed bias on the concatenated adaptation and evaluation \emph{raw
multi-channel EEG epochs}, applied after per-subject trunk pretraining
(calibration stays clean). Physically, this operator models
cross-talk between adjacent electrodes when conductive gel or saline
bridges form---a common field failure in long-duration dry-electrode
and gel-based montages---and cap shifts that rotate the spatial
projection of cortical sources relative to the montage. It is
\emph{not} a model of electrode dropout, impedance drift, or muscle
artifact; those require different operator families. Level \(0\) means
no synthetic stress; it is not necessarily distributionally identical
to calibration (especially Kumar).
\textbf{Scope of the positive recovery results:} all recovery and cost-bake-off
claims in Sections~\ref{sec:cost}--\ref{sec:skipl3} are made under this
\emph{controlled synthetic channel mixing}. Under natural session-to-session
drift (no synthetic operator) the factory trunk already transfers well, the
clean-referenced mean-shift gate does not open, and no recovery is claimed
there---we verify this directly with two revived natural-shift runs on
BNCI2014-004 (Section~\ref{sec:natural}). The honest framing is therefore: the
Personalizer recovers accuracy cheaply \emph{when deployment stress is strong
enough to break the factory}; on field-like drift its value is clean-safety and
cost, not accuracy.
\end{sloppypar}

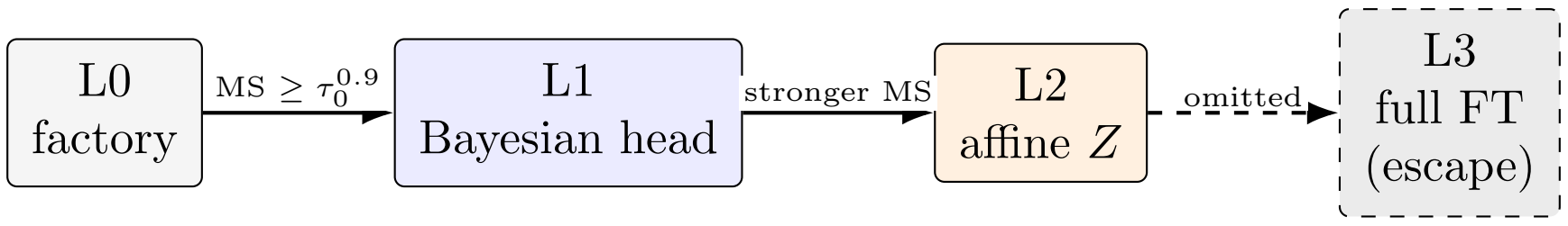
\begin{figure}[t]
\centering
\begin{tikzpicture}[
  box/.style={draw, rounded corners=2pt, align=center, inner sep=5pt,
              font=\small, minimum height=1.0cm},
  arr/.style={-{Latex}, thick},
  lbl/.style={above, font=\tiny, fill=white, inner sep=1pt}
]
\node[box, fill=black!4] (l0) {L0\\factory};
\node[box, fill=blue!8, right=1.4 of l0] (l1) {L1\\Bayesian head};
\node[box, fill=orange!12, right=1.4 of l1] (l2) {L2\\affine \(Z\)};
\node[box, fill=black!8, right=1.4 of l2, dashed] (l3) {L3\\full FT\\(escape)};
\draw[arr] (l0) -- node[lbl] {MS \(\ge \tau_0^{0.9}\)} (l1);
\draw[arr] (l1) -- node[lbl] {stronger MS} (l2);
\draw[arr, dashed] (l2) -- node[lbl] {omitted} (l3);
\end{tikzpicture}
\caption{Engineering ladder studied in this paper.
L0\(\to\)L1 uses a clean-referenced mean-shift gate.
L2 is the product ceiling under~\eqref{eq:U}; L3 is dashed because the
default policy omits it after comparing always-on baselines
(Section~\ref{sec:skipl3}).}
\label{fig:ladder}
\end{figure}

\paragraph{Default policy (clean-quantile cascade).}
\label{par:default}
The product default we study is a \emph{clean-quantile cascade}
(Figure~\ref{fig:ladder}):
\begin{enumerate}[leftmargin=1.4em,itemsep=0.15em]
\item score the stream with embedding mean-shift versus calibration;
\item enter L1 only if that score exceeds the leave-one-subject-out (LOSO) \(0.9\)-quantile of
scores on severity-\(0\) training folds (a clean-referenced gate);
\item allow escalation up to L2 under the same score family and~\eqref{eq:U};
\item omit L3 from the default policy (Section~\ref{sec:skipl3}), keeping
full fine-tune as an escape hatch.
\end{enumerate}
Higher thresholds remain allowed if they maximize utility subject to
\(\tau\ge\tau_{\mathrm{clean}}^{0.9}\).

\section{Cost: head versus encoder fine-tuning}
\label{sec:cost}

\begin{figure}[t]
\centering
\includegraphics[width=\linewidth]{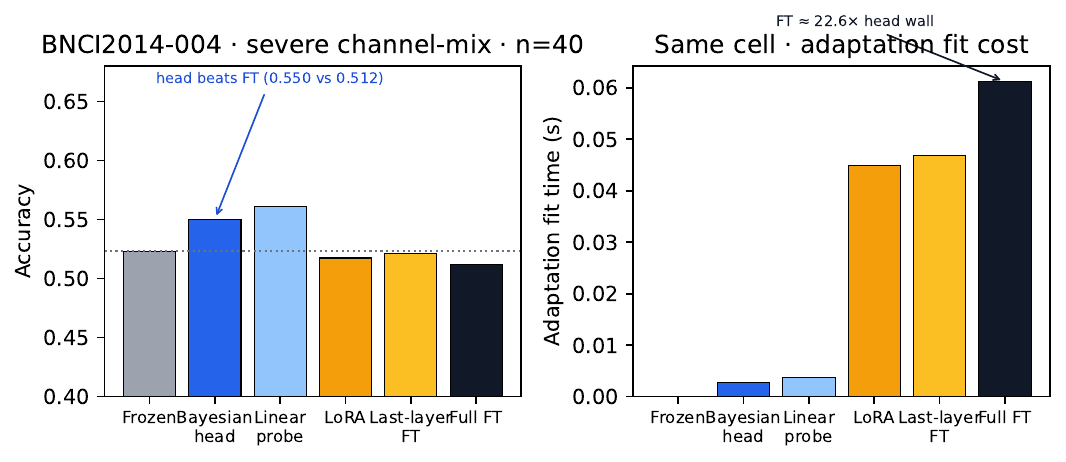}
\caption{BNCI2014-004 cost bake-off under severe channel-mix stress (\(n{=}40\)
labeled stream trials; mean over 9 subjects).
Left: accuracy (y-axis floor at \(0.4\) to show differences; absolute values
in Table~\ref{tab:cost}).
Right: adaptation fit time in our protocol.
The Bayesian head beats full fine-tune at a fraction of the fit cost.}
\label{fig:cost}
\end{figure}

Figure~\ref{fig:cost} and Table~\ref{tab:cost} report the BNCI cost
bake-off at channel-mix stress level \(0.75\), \(n{=}40\) (EEGNet trained on
calibration, then frozen; final evaluation metrics).
Relative to the frozen softmax (\(0.523\)), the LDA head reaches \(0.550\)
(\(\Delta\mathrm{acc}\approx +0.027\)) while full fine-tune falls to
\(0.512\) (below frozen), at adaptation fit time
\(\approx 0.003\,\mathrm{s}\) vs \(\approx 0.061\,\mathrm{s}\)
(\(\approx 23\times\)).
We therefore report absolute accuracy and wall ratios here rather than
``recovery of FT \(\Delta\mathrm{acc}\)'' (that ratio is not meaningful when
FT does not beat frozen).
Head expected calibration error (ECE~\citep{guo2017calibration}) is comparable to
full fine-tune on this cell; the linear probe is
slightly stronger on accuracy (\(0.561\)) at similar wall.
Short LoRA~\citep{hu2022lora} and last-layer fine-tune do not earn their
cost versus the head; both stay near the frozen baseline, and unsupervised
CORAL alignment~\citep{sun2016coral} falls below frozen.

\paragraph{Head-beats-frozen replicates across architectures.}
The headline cost cell above uses one trunk (EEGNet) and few subjects, so
its per-subject CI on head\,--\,FT includes zero (below). But the same LDA
head was run on five encoder trunks
(EEGNet/Shallow/Deep/Conformer/ATCNet) under the identical severe cell, and
that replication is significant on every dataset: pooled across the five
trunks, \(\Delta\mathrm{acc}(\mathrm{LDA}-\mathrm{frozen}) =
+0.065\), \([+0.027,+0.109]\) on BNCI (\(n{=}45\));
\(+0.086\), \([+0.043,+0.138]\) on Zhou (\(n{=}20\)); and
\(+0.030\), \([+0.011,+0.050]\) on Kumar (\(n{=}90\))---all CIs exclude
zero (2000 resamples).
Across the three datasets and five trunks together (\(n{=}155\)
subject\(\times\)trunk cells) the omnibus LDA\,--\,frozen effect is
\(+0.047\), \([+0.029,+0.065]\).
Because subjects recur across trunks, we also report a
\emph{cluster-aware} reanalysis (registered amendment; subjects
resampled as clusters with all their trunk results retained):
BNCI \(+0.065\), \([+0.004,+0.129]\); Zhou \(+0.086\),
\([+0.021,+0.170]\); Kumar \(+0.030\), \([+0.012,+0.048]\); omnibus
\(+0.047\), \([+0.025,+0.073]\) over 31 subject clusters. The
intervals widen as expected, and every one still excludes zero---the
replication survives the more conservative unit of inference.
(Mechanics: within-dataset intervals resample that dataset's subjects
with replacement; the omnibus resamples all 31 \((\)dataset,
subject\()\) clusters, so each \emph{subject}---not each cell---carries
equal weight across datasets. The registered amendment, the
reproduction gate, and the reanalysis script ship in the ancillary
bundle of this submission.)
These remain replications across \emph{architectures} on the same
subjects, not independent subject counts; the cross-architecture
consistency is itself the signal---the head's recovery does not hinge
on one trunk.

\begin{table}[t]
\centering
\caption{Apples-to-apples on frozen-trunk-\(Z\) (EEGNet): Personalizer head
family vs \(Z\)-space alternatives, severe channel-mix (severity \(0.75\),
\(n{=}40\)). BNCI: 9 subjects; Zhou: 4. Same cell as Table~\ref{tab:cost}
(heads/probe/coral/FT/frozen) plus QDA/Softmax heads on the same
embeddings. We omit a ``recovery vs FT'' column
because on BNCI EEGNet full fine-tune (\(0.512\)) falls \emph{below} frozen
(\(0.523\)), so that ratio is undefined there
(Section~\ref{sec:cost}). The frozen BNCI entry (0.523) is quoted from
the cost bake-off cell; the head bake-off CSV records 0.5215 for the
same arm. The source of the \(0.001\) difference is diagnosed to a
single trial: the two registered cells drew independent random
realizations of the synthetic mixing operator (same nine subjects,
same seed), and exactly one test trial flips across the realizations
(subject~4: \(0.70625\) vs \(0.69375\), the \(1/80\) accuracy
granularity) while the other eight subjects match exactly.}
\label{tab:external}
\small
\begin{tabular}{lcc}
\toprule
Arm (all on frozen-trunk-\(Z\)) & BNCI acc & Zhou acc \\
\midrule
Linear probe on \(Z\) & 0.561 & 0.525 \\
\textbf{QDA head} (Personalizer) & \textbf{0.558} & 0.500 \\
LDA head (Personalizer, default) & 0.550 & \textbf{0.540} \\
Softmax head (Personalizer) & 0.531 & \textbf{0.542} \\
Frozen EEGNet & 0.523 & 0.393 \\
Full fine-tune (L3) & 0.512 & 0.488 \\
CORAL+LDA (unsupervised) & 0.472 & 0.333 \\
\bottomrule
\end{tabular}
\end{table}

\paragraph{Single-trunk scope (the conservative cell).}
On the EEGNet cell specifically, the head's advantage over full fine-tune
is favourable on average but not significant at \(n{=}9\):
\(\Delta\mathrm{acc}(\mathrm{head}-\mathrm{FT}) = +0.038\),
95\% CI \([-0.007,0.081]\) on BNCI and \(+0.052\),
\([-0.023,0.147]\) on Zhou (\(n{\mathrm{subj}}{=}4\); pos.\ rate \(0.67\)
and \(0.50\)). On Zhou the frozen baseline is weaker (\(0.393\)), so the
LDA head reaches \(0.540\) (\(+0.147\)) and still beats full fine-tune
(\(0.488\)) at \(\sim\!22\times\) lower wall (linear probe close at
\(0.525\)). The cost contrast is thus the robust part of the single-cell
claim, and the multi-trunk replication above is the robust part of the
accuracy claim; parameter-efficient fine-tuning (PEFT) mid-rung options (LoRA, last-layer; our ``mid-hatches'') do not earn their
cost versus the head on either dataset. Fit-time ratios are
protocol-local and we do not merge them with other recovery setups.

\begin{table}[t]
\centering
\caption{BNCI2014-004 cost bake-off, channel-mix stress level \(0.75\),
\(n{=}40\) (mean over 9 subjects).
\(\Delta\) is vs frozen; wall ratio is FT wall / arm wall.}
\label{tab:cost}
\small
\begin{tabular}{lcccc}
\toprule
Arm & Acc & ECE & Fit time (s) & vs FT wall \\
\midrule
Frozen EEGNet & 0.523 & 0.223 & 0 & --- \\
\textbf{Bayesian head (LDA)} & \textbf{0.550} & 0.105 & \textbf{0.003} & \(\mathbf{\sim 23\times}\) \\
Linear probe & 0.561 & 0.074 & 0.004 & \(\sim 16\times\) \\
CORAL+LDA (unsup.) & 0.472 & 0.269 & 0.004 & \(\sim 16\times\) \\
LoRA & 0.517 & 0.126 & 0.045 & \(\sim 1.4\times\) \\
Last-layer FT & 0.521 & 0.182 & 0.047 & \(\sim 1.3\times\) \\
Full FT & 0.512 & 0.109 & 0.061 & \(1\times\) \\
\bottomrule
\end{tabular}
\end{table}

\subsection{Head family versus alternatives in a shared feature space}
\label{sec:external}

The Personalizer is a head over a \emph{frozen trunk's embeddings} \(Z\), so the
fair, apples-to-apples comparison is against other adaptation methods that
operate on the \emph{same} \(Z\)---not against pipelines on raw epochs, which
solve a different problem (Section~\ref{sec:crossfeat}). On the same severe
channel-mix cell (severity \(0.75\), \(n{=}40\)), we compare the Personalizer
head family (LDA/QDA/Softmax) against the alternatives that share its feature
space: a linear probe on \(Z\), unsupervised CORAL alignment~\citep{sun2016coral} on \(Z\) then LDA,
and full warm-start fine-tune of the trunk (Table~\ref{tab:external}).

On the refreshed EEGNet Z-bakeoff, every supervised head and the linear
probe exceed both frozen and full fine-tune on the BNCI severe cell, and
the LDA/Softmax/QDA heads likewise beat full fine-tune on Zhou. Full
fine-tune does not beat frozen on BNCI, so the honest claim is the weaker
one already in Table~\ref{tab:cost}: the LDA default beats FT on BNCI/Zhou
severe cells at \(\sim\!20\times\)--\(23\times\) lower wall, not that any
single head dominates by a large margin.
The Softmax head's bake-off utility is dominated by its fit wall time
(mean \(0.46\)~s in this protocol, \({\sim}100\times\) the conjugate
heads); read its accuracy row separately from the cascade economics.
The swappable head family remains the product story: LDA default, QDA as an
option when shift is severe. LDA remains the default over the
marginally stronger linear probe for a reason the accuracy column does
not price. Only the conjugate head supports \emph{exact online
updates}: its streaming update equals a batch refit bit-for-bit (a
registered benchmark equality), and it is stable at the small
calibration budgets deployment actually has. The probe is a
batch-only arm.

QDA is \emph{not} the best head on the EEGNet
severe cell (it is narrowly first on BNCI but last of the three on Zhou);
its ``best in \(13/20\) severe cells'' win is a \textbf{multi-trunk}
result (EEGNet/Shallow/Deep/Conformer/ATCNet on BNCI/Zhou/Shin/Kumar),
driven by the Deep/Conformer
trunks where QDA's covariance flexibility helps most---so QDA is an option,
not a universal default. Spending stream labels
on \(Z\) buys accuracy over label-free CORAL; the PEFT mid-hatches do not earn
their cost under~\eqref{eq:U}.

\paragraph{One boundary, three faces.}
Zhou remains the hard dataset for absolute factory accuracy (frozen
\(0.393\) on the refreshed bake-off),
so (a)~heads must be judged against a weak \(Z\); (b)~raw-epoch Riemannian
can still beat \(Z\)-space arms (Section~\ref{sec:crossfeat});
and (c)~the L3-omission recommendation's utility sensitivity is dataset-dependent
(Section~\ref{sec:usens}). The refreshed cost cell does \emph{not} keep
``FT beats head on Zhou'' (LDA head \(0.540\) vs FT \(0.488\)), but the
broader warning stands: when the deployed trunk's \(Z\) is weak, no head on
it can match a method that reaches richer features, and the right intervention
is feature-space selection,
not a better head. We therefore scope the Personalizer's accuracy advantage to
trunk\(\times\)dataset pairs where \(Z\) preserves class structure (BNCI here)
and report Zhou as the honest counter-cell.

\subsection{Cross-feature-space reference (raw / CSP pipelines)}
\label{sec:crossfeat}

For completeness we also ran classical pipelines that do \emph{not} use the
frozen trunk---Riemannian OAS+tangent+LogReg and
MDM~\citep{barachant2013classification}, Zanini-aligned MDM on raw
epochs, and CSP-feature LDA~\citep{ramoser2000excellent}---under the
identical split and budget. These
are not competitors of the Personalizer head (they require raw epochs, so they
apply when no frozen trunk is deployed); they sit in a different cell of the
design space and we report them as a reference, not a head-to-head. On BNCI the
best raw-epoch arm is CSP-feature LDA (\(0.609\)), above the Personalizer's
\(Z\)-space arms (QDA \(0.558\), linear probe \(0.561\)); the Riemannian
tangent (\(0.577\)) is on par with them. On Zhou the Riemannian
raw-epoch pipeline (\(0.613\)) is likewise
stronger than any \(Z\)-space arm. The pattern is consistent across both
datasets: when raw epochs are available, a classical pipeline can match or
exceed what a head on frozen \(Z\) reaches---the same ``weak \(Z\)'' boundary
discussed above, seen here from the raw-feature side.

\subsection{Natural-drift boundary: where the recovery story stops}
\label{sec:natural}

The recovery wins above all sit under controlled synthetic channel mixing. A
fair scope check asks what happens under \emph{natural} session-to-session
drift, with no synthetic operator. We revived two BNCI2014-004 scenarios for
this (9 subjects, no perturbation): a same-day session-recovery split and a
sparse-factory cross-day split. In both the factory trunk already transfers
well and the clean-referenced mean-shift gate \textbf{does not open}
(an absolute accuracy-degradation rule at \(<0.70\) likewise fails): frozen
means \(0.735\) and \(0.761\) respectively. With no factory degradation to
recover from, the Personalizer head shows no reliable accuracy gain over
frozen at \(n{=}20\)
(\(\Delta\mathrm{acc}\) \(+0.025\) on session-recovery and \(-0.012\) on
sparse-cross; both subject CIs include zero). Only the cost contrast survives
there (adapt-fit wall far below warm-start fine-tune). This is the boundary we
claim honestly: the Personalizer's accuracy advantage is real \emph{when stress
breaks the factory}; on field-like drift the advantage is cost and clean-safety,
not accuracy.

\paragraph{Reconciling with the drift table.}
Table~\ref{tab:drift} reports a positive mean head benefit under field-like
drift at \(n{=}40\) (BNCI \(+0.056\), Kumar \(+0.030\)), which is not the same
as a reliable per-subject recovery claim. The two numbers are consistent:
Table~\ref{tab:drift} pools a different split (three-session S3, oracle
\(\Delta\mathrm{acc}\) on the held-out evaluation session) and its
oracle-fire rates are only \(0.56\) (BNCI) and \(0.61\) (Kumar)---i.e.\ the
head helps on roughly half the subjects, so the positive mean is driven by a
subset of subjects and does not contradict the same-day protocol above, where
subject CIs include zero. The honest summary is that under natural drift the
head is sometimes oracle-useful but the gate cannot identify those subjects
from clean alone, which is exactly why a utility-max LOSO gate over-adapts
(Section~\ref{sec:cq}) and a clean-referenced gate is needed.

\section{Ladder and regime-aware gating}
\label{sec:ladder}

\subsection{Why utility-max LOSO fails on field-like clean}
\label{sec:cq}

A natural score for leaving L0 is embedding mean-shift between calibration
and stream.
Fitting a LOSO threshold by maximizing~\eqref{eq:U} overfires on clean
cells that already carry session drift: on BNCI2014-004 severity \(0\)
every clean subject adapts (rate \(1.00\); bootstrap 95\% CI
\([1.00,1.00]\), \(n{=}9\)), and on Kumar2024 severity \(0\) the rate is
still \(0.56\) (\([0.33,0.78]\), \(n{=}18\))---both well above the
\(\approx\!0\) a clean-referenced gate achieves.

Severity \(0\) on Kumar is not factory-clean.
At \(n{=}40\), the rate at which an oracle prefers the head over the frozen
decoder is \(0.61\) and mean
\(\Delta\mathrm{acc}(\mathrm{head}-\mathrm{frozen})\approx +0.030\),
versus BNCI \(0.56\) / \(+0.056\)
(Table~\ref{tab:drift}).
Mean-shift scores are already elevated, and utility-max LOSO lowers the
threshold further because the head often helps.

\begin{table}[t]
\centering
\caption{Natural session drift at severity \(0\), \(n{=}40\)
(no synthetic shift; Mahalanobis mean-shift default).
``Oracle fire'' = fraction of subjects where head beats frozen on evaluation
accuracy.}
\label{tab:drift}
\small
\begin{tabular}{lcccc}
\toprule
Dataset & \(n\) & Oracle fire & \(\Delta\)acc head$-$frozen & Mean MS \\
\midrule
BNCI2014-004 & 9 & 0.56 & \(+0.056\) & 0.63 \\
Zhou2016 & 4 & 0.25 & \(-0.027\) & 0.69 \\
Kumar2024 & 18 & 0.61 & \(+0.030\) & 0.62 \\
\bottomrule
\end{tabular}
\end{table}

\subsection{Clean-quantile gate}
\label{sec:cqgate}

We replace the L0\(\to\)L1 cut by a \emph{clean-referenced} threshold:
LOSO \(\tau_0 =\) the \(0.9\)-quantile of mean-shift on severity-\(0\)
training folds.
Intuitively: adapt only when shift exceeds ordinary cross-session noise.
The clean-quantile cascade of paragraph~\ref{par:default} uses this gate
and caps escalation at L2.

\begin{table}[t]
\centering
\caption{Clean (no synthetic shift) safety at \(n{=}40\).
Trigger adapt-rate rows are subject-level (\(n{=}9\) BNCI, \(n{=}18\)
Kumar); cascade factory-stay rows pool subject\(\times\)trunk cells
(5 trunks). CIs are paired bootstrap (2000 resamples).}
\label{tab:cq}
\small
\begin{tabular}{llcc}
\toprule
Dataset & Gate / policy & Clean rate & 95\% CI \\
\midrule
BNCI & mean-shift (utility-max) & adapt \(1.00\) & \([1.00,1.00]\) \\
Kumar & mean-shift (utility-max) & adapt \(0.56\) & \([0.33,0.78]\) \\
Kumar & \textbf{mean-shift (clean-quantile)} & adapt \(\mathbf{0.11}\) & \([0.00,0.28]\) \\
Kumar & cascade max-L2 (no clean ref.) & factory \(0.51\) & \([0.41,0.61]\) \\
Kumar & \textbf{clean-quantile cascade} & factory \(\mathbf{0.87}\) & \([0.79,0.93]\) \\
\midrule
BNCI & clean-quantile cascade & factory \(0.87\) & \([0.76,0.96]\) \\
Zhou & clean-quantile cascade & factory \(0.70\) & \([0.50,0.90]\) \\
\bottomrule
\end{tabular}
\end{table}

Table~\ref{tab:cq} shows the clean-safety win
(final evaluation data).
As a deployment-oriented summary: the utility-max mean-shift gate
over-adapts on clean BNCI (\(1.00\)) and clean Kumar (\(0.56\)); the
clean-quantile \emph{trigger} drops Kumar to \(11\%\), and the cq
\emph{cascade} stays on the factory for \(\approx\!87\%\) of
clean cells (Kumar/BNCI) and \(70\%\) (Zhou).
On severe BNCI/Zhou, the clean-quantile cascade is comparable to (not
significantly above) an always-on head policy on utility
(paired \(\Delta U\): BNCI \(+0.010\), CI \([-0.006,0.028]\); Zhou
\(+0.013\), CI \([-0.003,0.032]\); both CIs include zero).
On Kumar severe, the cascade-versus-always-on-head gap is
\(\Delta U\approx -0.005\), CI \([-0.013,0.003]\): the gate restores
clean factory use; it does \emph{not} create affine recovery where none
exists.

\paragraph{No statistically detectable gate cost on severe.}
The clean-reference threshold is a \emph{false-entry control}, not a
detector of when adaptation will help: oracle fire rates (the fraction of
subjects where the head beats frozen) are near chance
(Section~\ref{sec:cq}) mean the gate cannot identify which subjects
benefit---it controls wasted entry, not benefit prediction. With that
scope fixed, we ask whether gating costs accuracy:
the CIs below include negative values, and a non-significant
difference is not evidence of equivalence---the claim is \emph{no
detectable loss in these samples}, not non-inferiority.
Capping the cascade at L2 risks losing accuracy relative to an always-on
head on severe cells; we check the paired subject \(\Delta\mathrm{acc}\)
(cascade \(-\) always-head, eegnet, severe \(n{=}40\), 2000 resamples):
BNCI \(+0.024\), \([-0.003,0.054]\) (favourable on average but not
significant; pos.\ rate \(0.44\)); Zhou \(+0.025\), \([+0.003,0.057]\)
(CI excludes zero); Kumar \(0.000\), \([0.000,0.000]\).
The cascade shows \emph{no statistically detectable} accuracy loss
relative to always-head on severe in these samples; point estimates
are at or above zero, though the BNCI interval includes negative
values. Meanwhile the cascade stays on the factory for
\(\approx\!87\%\) of clean cells where always-head would waste
labels. (On Zhou the cascade is significantly \emph{above}
always-head; on BNCI and Kumar it is on par.) The deployment trade
therefore reads one-sided \emph{in these data}: no detectable
severe-cell cost, large clean-safety gain (\(0.87\) vs \(0.51\)
factory-stay without the clean-quantile gate) at \(\sim\!23\times\)
lower fit wall.

\subsection{Why the default omits full fine-tune}
\label{sec:skipl3}

\begin{figure}[t]
\centering
\includegraphics[width=\linewidth]{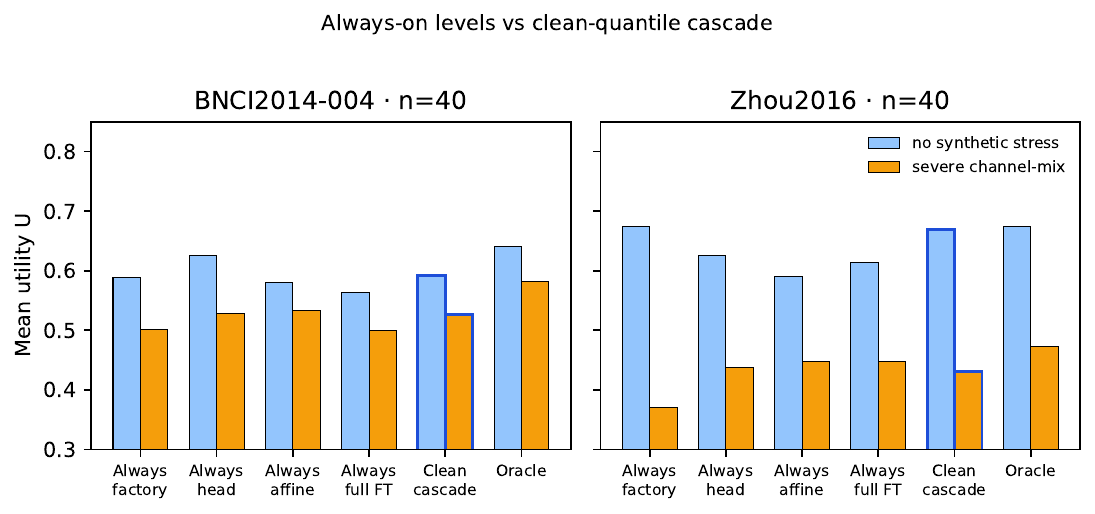}
\caption{Always-on levels versus the clean-quantile cascade and an oracle
level selector at \(n{=}40\).
Blue: no synthetic shift; orange: severe channel-mix stress.
The cascade stays near factory utility when clean and near always-on
affine utility under severe shift, without always paying full fine-tune.}
\label{fig:baselines}
\end{figure}

We do not omit L3 merely by assertion.
First we compare \emph{always-on} level policies under~\eqref{eq:U}
(always factory, always head, always affine, always full fine-tune) on the
focus cells against the clean-quantile cascade and an oracle level selector
(Figure~\ref{fig:baselines}).
On BNCI focus cells, always-on full fine-tune is not the mean-utility
winner among these baselines: pricing fit time and labels makes full
fine-tune rarely optimal even when it raises accuracy.
On Zhou, always-on full fine-tune can be competitive among always-on
baselines on some mid-severity cells, but under severe shift the cascade
still tracks near always-on affine without paying always-on fine-tune
cost on clean.
\textbf{Therefore} the product default uses a design ceiling at L2 (the
clean-quantile cascade never selects L3), while full fine-tune remains an
escape hatch outside the default policy.
This is a \textbf{product default under fixed \(U\)}, not a claim that
full fine-tune is never useful under any utility or cell.

\subsection{Utility-weight sensitivity of the L3-omission recommendation}
\label{sec:usens}

\begin{figure}[t]
\centering
\includegraphics[width=\linewidth]{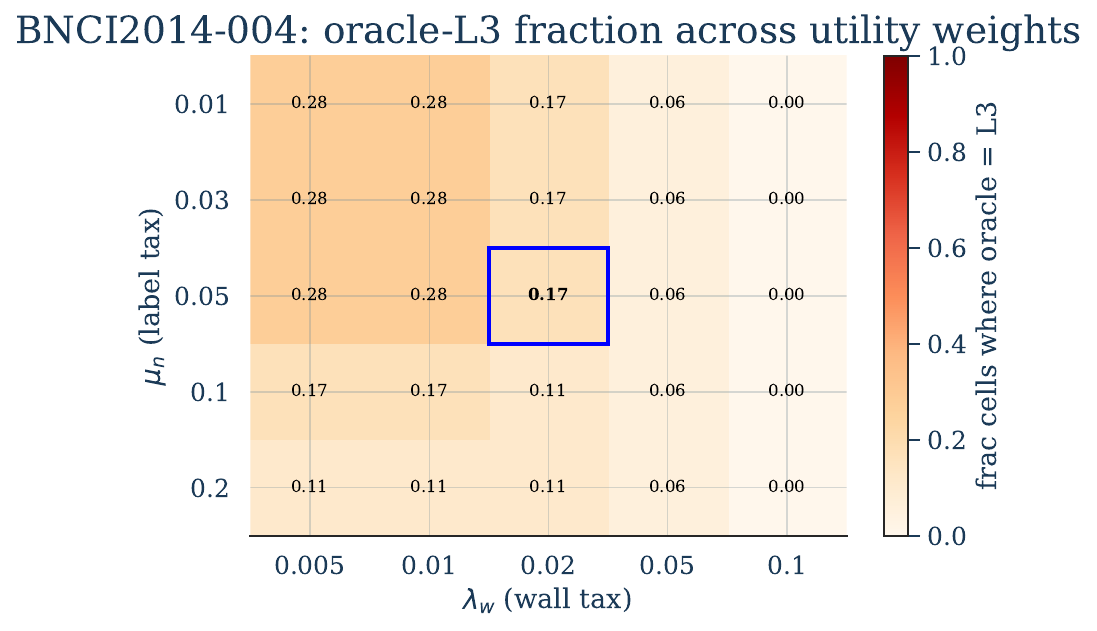}
\caption{Fraction of cells where full fine-tune (L3) is the oracle utility
winner, across the \((\lambda_w, \mu_n)\) grid (BNCI2014-004; no models
re-run---per-level utilities re-derived from the evaluation data).
The chosen default \((0.02, 0.05)\) is boxed (\(0.17\)). Companion heat-maps for
Zhou/Kumar are in the ancillary bundle; fixed-weight oracle-L3 fractions are
\(0.25\) / \(0.22\).}
\label{fig:usens}
\end{figure}

A natural objection to the chosen default is that the two weights in~\eqref{eq:U}
are free, and pricing wall-time plus labels is exactly what makes the expensive
option (L3) lose. We test whether omitting L3 is a point artifact or a robust
region by re-deriving per-level utilities from the existing evaluation data under
a grid \(\lambda_w \in \{0.005,0.01,0.02,0.05,0.1\} \times
\mu_n \in \{0.01,0.03,0.05,0.1,0.2\}\)---without re-running any model---and
measuring the fraction of (subject\(\times\)severity\(\times\)budget) cells
where L3 is the oracle utility winner (Figure~\ref{fig:usens}).

Re-derived utilities (no models re-run):
at the fixed weights, the oracle-L3 fraction is \(0.17\) (BNCI), \(0.22\)
(Kumar), and \(0.25\) (Zhou); across the full grid the ranges are
roughly \(0\)--\(0.28\), \(0.03\)--\(0.39\), and \(0\)--\(0.50\).
The clean-quantile cascade also beats always-L3 on mean utility at the
fixed weights on all three datasets.
We therefore treat the L3 omission as \textbf{robust on BNCI/Kumar} and
\textbf{still preferred on Zhou at the fixed weights}, with the caveat that
Zhou's grid max reaches \(0.50\) when labels/wall are priced very cheaply---so
Zhou remains the most weight-sensitive cell, not a fixed-weight inversion.

\section{No universal adaptation policy}
\label{sec:transfer}

\begin{figure}[t]
\centering
\includegraphics[width=0.72\linewidth]{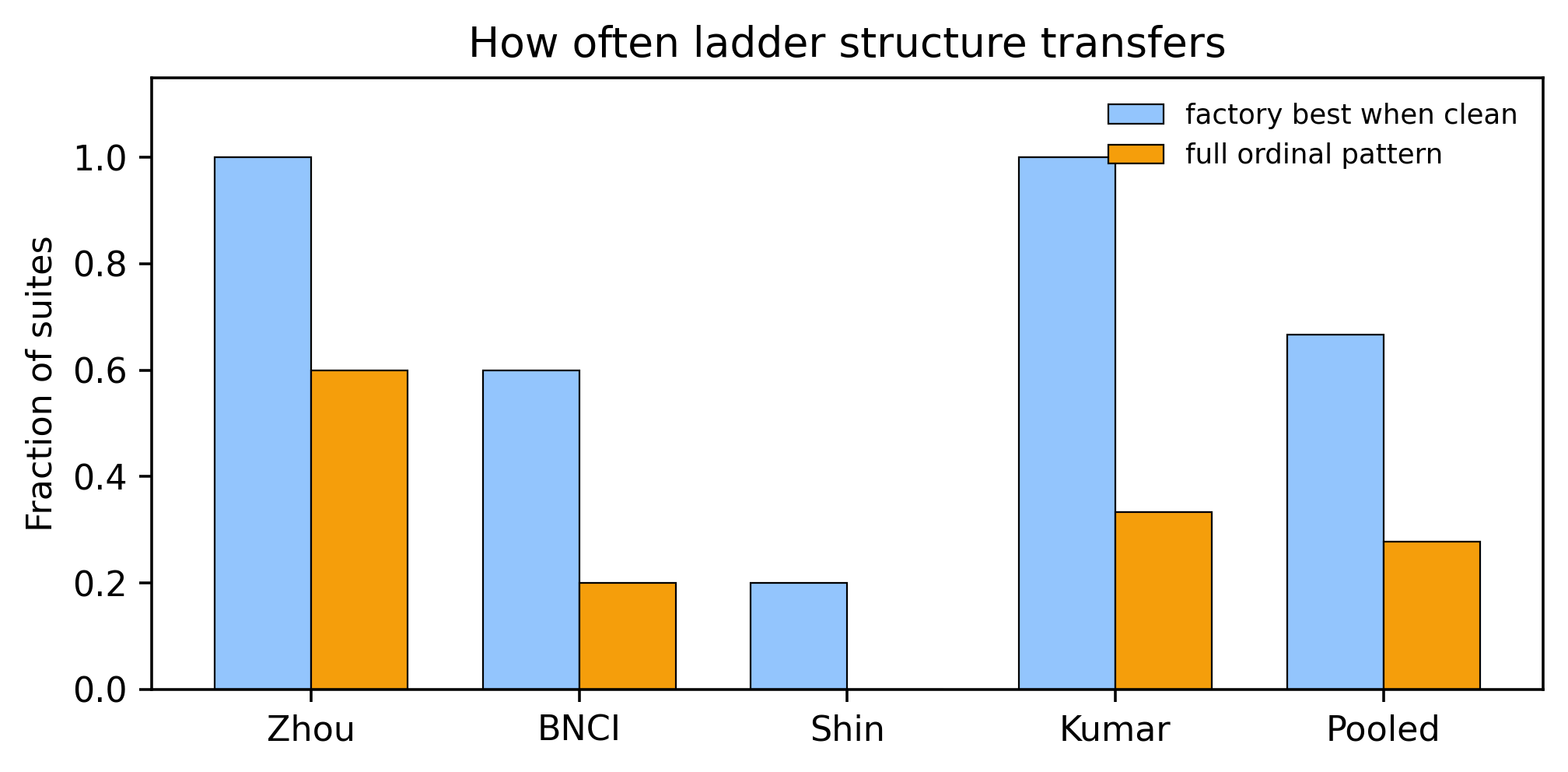}
\caption{Per-suite policy hold rates (same counts as
Table~\ref{tab:transfer-cells}): clean\(\to\)L0 holds in 12/18
trunk\(\times\)dataset cells; the full ordinal ladder in only 5/18.
Zhou is the easy suite; Shin carries neither.}
\label{fig:transfer}
\end{figure}

The gating and cost results do not imply one ordinal ladder for every
trunk and dataset.
\textbf{There is no universal adaptation policy} in our suite map
(Figure~\ref{fig:transfer}).

What transfers most reliably is soft:
when there is no synthetic shift, the factory decoder is often best
(\(12/18\) trunk\(\times\)dataset cells; Table~\ref{tab:transfer-cells}).
That matches the clean-quantile gate's job---do not always adapt on clean.

What does \emph{not} transfer universally is a stricter ordinal pattern
(clean best = L0, severe best = L2, and positive severity--level
correlation): only \(5/18\) suites.
We report that fraction as a \textbf{boundary map}, not as a success rate:
severity\(\to\)adaptation-level does not copy across architectures and
datasets even when factory-when-clean sometimes does.

\paragraph{Statistical caveat (counts are small).}
Both pooled fractions carry substantial uncertainty. A Wilson 95\% CI on the
\(12/18\) clean-transfer rate is \([0.44, 0.84]\): the interval includes the
\(0.5\) chance rate of ``any one level wins,'' and against that null the
one-sided \(p=0.119\) does not survive Holm/BH correction across the
count-claim family.
The strict-ordinal \(5/18\) is \(0.28\), Wilson \([0.12, 0.51]\), not
significant against the \(0.25\) null for two independent conditions
(\(p=0.481\)).
We therefore frame the transfer claim as a \emph{soft, dataset-dependent}
tendency---strong on Zhou, weak on Shin---not a quantified universal law.

We evaluate on encoder trunks
\{EEGNet, Shallow, Deep, Conformer, ATCNet\}
\(\times\) datasets BNCI, Zhou, Shin (budget clamped when the stream is
short), and EEGNet/Shallow/Conformer on Kumar.

Zhou supports clean\(\to\)L0 on all five trunks and the full ordinal pattern
on three (5/5 clean, 3/5 ordinal); BNCI is mixed (3/5 clean, 1/5 ordinal);
Shin fails under short streams (1/5 clean, 0/5 ordinal); Kumar is
clean-safe but rarely ordinal (3/3 clean, 1/3 ordinal).
The actionable soft principle is: \emph{do not always adapt on clean}.

\begin{table}[t]
\centering
\caption{Per-suite boundary map. A check (\(\checkmark\)) means the principle
holds for that trunk\(\times\)dataset cell; a dot means it fails. The
pattern is dataset-driven, not trunk-driven: Zhou carries the ordinal
story, Shin (short streams) carries none, Kumar is clean-safe but rarely
ordinal, and BNCI is mixed. This is why we report a boundary map, not a
universal policy.}
\label{tab:transfer-cells}
\small
\begin{tabular}{llccc}
\toprule
Dataset & Trunk & clean\(\to\)L0 & severe\(\to\)L2 & full ordinal \\
\midrule
Zhou2016        & ATCNet    & \(\checkmark\) & \(\checkmark\) & \(\checkmark\) \\
Zhou2016        & Conformer & \(\checkmark\) & \(\checkmark\) & \(\checkmark\) \\
Zhou2016        & EEGNet    & \(\checkmark\) & \(\checkmark\) & \(\checkmark\) \\
Zhou2016        & Deep      & \(\checkmark\) & $\cdot$ & $\cdot$ \\
Zhou2016        & Shallow   & \(\checkmark\) & $\cdot$ & $\cdot$ \\
BNCI2014-004    & ATCNet    & \(\checkmark\) & $\cdot$ & $\cdot$ \\
BNCI2014-004    & Conformer & \(\checkmark\) & \(\checkmark\) & \(\checkmark\) \\
BNCI2014-004    & Deep      & \(\checkmark\) & $\cdot$ & $\cdot$ \\
BNCI2014-004    & EEGNet    & $\cdot$ & \(\checkmark\) & $\cdot$ \\
BNCI2014-004    & Shallow   & $\cdot$ & \(\checkmark\) & $\cdot$ \\
Shin2017A       & EEGNet    & \(\checkmark\) & $\cdot$ & $\cdot$ \\
Shin2017A       & ATCNet    & $\cdot$ & \(\checkmark\) & $\cdot$ \\
Shin2017A       & Conformer & $\cdot$ & \(\checkmark\) & $\cdot$ \\
Shin2017A       & Deep      & $\cdot$ & $\cdot$ & $\cdot$ \\
Shin2017A       & Shallow   & $\cdot$ & $\cdot$ & $\cdot$ \\
Kumar2024       & EEGNet    & \(\checkmark\) & $\cdot$ & $\cdot$ \\
Kumar2024       & Shallow   & \(\checkmark\) & \(\checkmark\) & \(\checkmark\) \\
Kumar2024       & Conformer & \(\checkmark\) & $\cdot$ & $\cdot$ \\
\midrule
\textbf{Pooled} &           & \textbf{12/18} & \textbf{9/18} & \textbf{5/18} \\
\bottomrule
\end{tabular}
\end{table}

\section{Discussion}
\label{sec:discussion}

\paragraph{Product surface.}
The deployment path this paper supports is a control layer over a frozen
encoder: monitor embedding shift, decide whether to spend adaptation
resources, escalate a lightweight ladder (head, affine), and keep full
fine-tune as an escape hatch. The companion Personalizer
paper~\citep{musienko2026personalizer} ships the trunk-agnostic API that
realizes this; here we supply the gate and ladder that govern it.
Under~\eqref{eq:U}, the chosen default is the clean-quantile cascade of
paragraph~\ref{par:default}.
Richer automatic choice among mid-tier recovery mechanisms is out of
scope here.

\paragraph{Scope.}
This paper is about resource-aware control and gating---not a new decoder
and not foresight theory.
Synthetic channel mixing stresses the cost bake-off; natural session drift
stresses the gate.
The ladder and cost figures therefore use \textbf{one} synthetic
operator family by design.
Supporting same-day BNCI2014-004 EEGNet recovery cells under a different
protocol (LOSO factory, cal/eval on one day, \(n{=}20\)) also show cheap
Personalizer recovery for \textbf{one-channel dropout} and mild channel
mixing (Table~\ref{tab:other-stress}).
Treat that table as an appendix-style second synthetic family---not primary
evidence, not a remapping of the ladder severity grid, and not physical
electrode or clinical SNR evidence.
Fit-wall ratios in that harness (\({\sim}30\times\)) must not be quoted as
the bake-off cost story and must not be merged with the protocol-local walls
in Section~\ref{sec:cost}.
The evidence is MI-only by design: the ladder prices adaptation
resources on the paradigm where labelled shift is cheapest to study,
and the companion trust line extends the shipped surface to
SSVEP~\citep[p.~16]{musienko2026personalizer}; replicating the ladder
itself on further paradigms is future work.

\begin{table}[t]
\centering
\caption{Exploratory supporting cells only (different protocol than the
ladder bake-off; not primary evidence).
BNCI2014-004 EEGNet, \(n{=}20\).
\(\Delta\mathrm{acc}\) is Personalizer minus frozen;
wall ratio is fine-tune / Personalizer adapt fit.}
\label{tab:other-stress}
\small
\begin{tabular}{lcccc}
\toprule
Synthetic stress & Frozen acc & Pers.\ acc & \(\Delta\mathrm{acc}\) & Wall FT/Pers. \\
\midrule
1-channel dropout & 0.654 & 0.740 & \(+0.086\) & \({\sim}30\times\) \\
Mild channel mix & 0.542 & 0.756 & \(+0.214\) & \({\sim}34\times\) \\
\bottomrule
\end{tabular}
\end{table}

\FloatBarrier
\paragraph{Utility sensitivity.}
Weights in~\eqref{eq:U} are fixed for comparability; they encode one
cued-deployment scenario. At these weights, one accuracy point
(\(0.01\)) trades against \(0.5\)~s of protocol-local fit wall or 16
stream labels (fit wall excludes data transfer and preprocessing).
Omitting L3 is not a point artifact: across the
\((\lambda_w,\mu_n)\) sweep (Section~\ref{sec:usens}) the
fixed-weight oracle-L3 fraction is \(0.17\)/\(0.22\)/\(0.25\) on
BNCI/Kumar/Zhou, and the clean-quantile (cq) cascade beats always-L3 on
mean \(U\) at those weights on all three. Zhou remains the most
weight-sensitive cell (grid max \(0.50\) when labels/wall are cheap).
\FloatBarrier

\paragraph{Beyond deciding to adapt: certifying when to act.}
\label{sec:forward}
This paper prices the decision to spend adaptation resources. The
complementary half of deployment trust is deciding \emph{when to act} on
the adapted decoder's output. The same research
programme~\citep{romano2020classification,angelopoulos2023gentle} ships
split-conformal prediction-set gating over the adapted posterior:
distribution-free bounds on the wrong-command rate under exchangeability,
and a certificate that automatically suspends itself whenever the head
adapts. The evaluation runs in preregistered benchmark cells on three
public MI corpora
and an SSVEP speller corpus. We keep it out of scope here by design:
gating adaptation and certifying decisions are separable layers, and
conflating them would blur both claims.

\paragraph{Scope, statistics, and limitations.}
All contrasts are exploratory subject-level bootstrap summaries; the
subject counts (BNCI 9, Zhou 4, Shin 9, Kumar 18) are the full cohort
sizes of these public corpora, not a subsample. Cross-architecture
replication spans five encoder trunks; because subjects are shared
across trunks, the cluster-aware reanalysis (Section~\ref{sec:cost};
31 subject clusters) provides the conservative intervals, and all
headline directions survive it. Shin streams are short; Kumar online
sessions differ from offline MI; fit times are protocol-local. Labels
come from the cued paradigm; in deployment, label provenance is itself
a cost our utility prices but does not solve---active-learning stopping
and error-potential labeling are the shipped product's answers, out of
scope here. Every table is tied to its registered CSV by a
content-hash digest in the evidence manifest (shipped in the ancillary
bundle).

\section{Conclusion}
\label{sec:conclusion}

A deployed BCI decoder should spend adaptation resources under a gate, not
by default.
A clean-referenced mean-shift rule prevents false entry into head
adaptation when ``clean'' already carries session drift (illustrated on
BNCI and Kumar online sessions).
When adaptation is warranted, a cheap LDA head recovers accuracy under
severe \emph{controlled channel-mixing} stress and replicates across
five encoder architectures (pooled
LDA\,--\,frozen \(+0.047\), \([+0.025,+0.073]\) cluster-aware over 31
subjects; \(n{=}155\) subject\(\times\)trunk cells), at
\(\sim\!23\times\) lower fit time than full fine-tune---the cheaper rung
on a gated ladder, not a universal replacement for fine-tuning.
Under a shared utility, always-on baselines motivate capping the product
default below full fine-tune; the PEFT mid-hatches do not earn cascade rungs.
There is no universal ordinal adaptation policy across trunks and
datasets (\(5/18\)); factory-when-clean is the more transferable soft rule
(\(12/18\)).
That is the control-layer story: decide first, then adapt cheaply.

\section*{Availability}
The control layer described here ships in the \texttt{nimbus-bci} Python
package (v0.9.0, \texttt{pip install nimbus-bci}): the ladder arms are
the \texttt{Personalizer} head family, and the clean-quantile cascade
with utility capping ships as the factory decision presets. The
evidence manifest, the frozen CSVs behind every table, and the
cluster-bootstrap reanalysis script ship as ancillary files with this
arXiv submission---retrievable without any account or private access.
Documentation:
\url{https://docs.nimbusbci.com}. The companion Personalizer API paper
is available as arXiv:2607.22397~\citep{musienko2026personalizer}.

\section*{Acknowledgments}
Exploratory results were produced with the preregistered
\texttt{nimbusbench} suite accompanying Nimbus Personalizer; the suite
is not required to use the Personalizer API.

\appendix

\section{Mean-shift statistic and conjugate head update}
\label{app:formulas}

This appendix makes the paper self-contained: the two mathematical
objects referenced in the body but deferred to the companion are stated
here explicitly.

\paragraph{Embedding mean-shift (the gate's score).}
Let \(Z_{\mathrm{cal}} = \{z_i\}_{i=1}^{n}\) be the calibration
embeddings and \(Z_{\mathrm{stream}}\) the current stream window. The
mean-shift score used by the L0\(\to\)L1 gate is the (squared)
Mahalanobis distance between the two means, under the calibration
covariance:
\begin{equation}
\mathrm{MS}(Z_{\mathrm{cal}}, Z_{\mathrm{stream}})
= \bigl(\bar{z}_{\mathrm{stream}} - \bar{z}_{\mathrm{cal}}\bigr)^{\!\top}
\Sigma_{\mathrm{cal}}^{-1}
\bigl(\bar{z}_{\mathrm{stream}} - \bar{z}_{\mathrm{cal}}\bigr),
\label{eq:ms}
\end{equation}
where \(\bar{z}\) is the arithmetic mean and
\(\Sigma_{\mathrm{cal}}\) is the (regularized) covariance of the
calibration embeddings. The clean-quantile threshold
\(\tau_0^{0.9}\) is the 0.9-quantile of~\eqref{eq:ms} evaluated on
LOSO severity-0 calibration folds (Section~\ref{sec:cqgate}).

\paragraph{Conjugate head (NIW prior, closed-form update).}
The L1 head is a Bayesian Gaussian classifier with a shared covariance.
For each class \(k \in \{1, \dots, K\}\):
\begin{align}
z_i \mid y_i = k &\;\sim\; \mathcal{N}(\mu_k, \Sigma), \label{eq:lik}\\
\mu_k, \Sigma &\;\sim\; \mathrm{NIW}(\mu_{0,k}, \kappa_0, \Psi, \nu_0),
\label{eq:prior}
\end{align}
where \(\mathrm{NIW}\) is the Normal--Inverse--Wishart conjugate
prior with location \(\mu_{0,k}\), shrinkage \(\kappa_0\), scale
\(\Psi\), and degrees of freedom \(\nu_0\). Given a batch of labelled
embeddings \(\{(z_i, y_i)\}_{i=1}^{n}\), the posterior is again NIW
with parameters updated in closed form:
\begin{align}
\kappa_n &= \kappa_0 + n_k, &
\mu_{n,k} &= \frac{\kappa_0 \mu_{0,k} + n_k \bar{z}_k}{\kappa_n},
\label{eq:post-mu}\\
\nu_n &= \nu_0 + n, &
\Psi_n &= \Psi + S + \sum_{k=1}^{K}
\frac{\kappa_0 n_k}{\kappa_n}(\bar{z}_k - \mu_{0,k})(\bar{z}_k - \mu_{0,k})^{\!\top},
\label{eq:post-Sigma}
\end{align}
where \(n_k\) is the count of class-\(k\) trials, \(\bar{z}_k\) their
mean, and \(S = \sum_i (z_i - \bar{z}_{y_i})(z_i - \bar{z}_{y_i})^{\!\top}\)
the within-class scatter. The posterior predictive for a new embedding
\(z_{\ast}\) reduces to a Linear Discriminant (LDA) decision rule with
plug-in parameters from~\eqref{eq:post-mu}--\eqref{eq:post-Sigma};
the streaming update (\(\texttt{partial\_fit}\)) applies
equations~\eqref{eq:post-mu}--\eqref{eq:post-Sigma} incrementally and
equals a batch refit exactly (a bit-level equality registered in the
benchmark suite; Section~\ref{sec:cost}). Full regularization
constants and the QDA/Softmax variants are in the companion
paper~\citep{musienko2026personalizer}.

\bibliographystyle{plainnat}
\bibliography{references}

\end{document}